%% file: main.tex
\documentclass{article}

\usepackage{spconf}
\usepackage{amsmath}
\usepackage{amsfonts}
\usepackage{amssymb}
\usepackage{amsthm}
\usepackage{graphicx}
\usepackage[numbers]{natbib}
\usepackage{siunitx}
\usepackage{hyperref}
\usepackage{booktabs}

\usepackage[dvipsnames]{xcolor}
\usepackage[skip=6pt]{caption}

\newcommand\myshade{70}
\colorlet{mywholecolor}{MidnightBlue}
\hypersetup{
  linkcolor  = mywholecolor!\myshade!black,
  citecolor  = mywholecolor!\myshade!black,
  urlcolor   = mywholecolor!\myshade!black,
  colorlinks = true,
}

\title{Modeling Analog Dynamic Range Compressors \\ using Deep Learning and State-space Models}

\name{
    Hanzhi Yin$^{1,2}$,
    Gang Cheng$^{2}$,
    Christian J. Steinmetz$^{3}$,
    Ruibin Yuan$^{2}$,
    Richard M. Stern$^{1}$,
    Roger B. Dannenberg$^{1}$
}

\address{
    $^{1}$ School of Music, Carnegie Mellon University \\
    $^{2}$ Multimodal Art Projection Research Community \\
    $^{3}$ Centre for Digital Music, Queen Mary University of London \\
}

\begin{document}

\maketitle

\begin{abstract}

We describe a novel approach for developing realistic digital models of dynamic range compressors for digital audio production by analyzing their analog prototypes.
While realistic digital dynamic compressors are potentially useful for many applications, the design process is challenging because the compressors operate nonlinearly over long time scales.
Our approach is based on the structured state space sequence model (S4), as implementing the state-space model (SSM) has proven to be efficient at learning long-range dependencies and is promising for modeling dynamic range compressors.
We present in this paper a deep learning model with S4 layers to model the Teletronix LA-2A analog dynamic range compressor.
The model is causal, executes efficiently in real time, and achieves roughly the same quality as previous deep-learning models but with fewer parameters.

\end{abstract}


\section{Introduction} \label{sec:intro}

Virtual analog modeling (VA modeling) concerns the digital simulation of analog audio devices like synthesizers and audio effect units~\cite{Pekonen2011TheSynthesis, Werner2016VirtualFilters}.
There has been a trend of using deep learning (DL) techniques in VA modeling, transforming VA modeling tasks to be data-driven by utilizing input and output waveform pairs processed by the analog system.
Introducing VA modeling to DL may bring some research benefits.
While these approaches can be used to emulate analog audio systems, they can also be used to construct differentiable proxies~\cite{Engel2020DDSP:Processing}, facilitating tasks such as automatic mixing~\cite{Steinmetz2021AutomaticEffects}.

Until now, applications of DL to VA modeling have focused mostly on vacuum-tube amplifiers~\cite{Wright2019Real-timeNetworks, Damskagg2019DeepEmulation} and distortion pedals~\cite{Wright2020Real-TimeLearning, Damskagg2019Real-timeLearning}.
In contrast, dynamic range compressors (DRCs), which are non-linear, time-invariant, and possess longer temporal dependencies, have received less attention.
Existing attempts include the use of autoencoders~\cite{Hawley2019SignalTrain:Networks}, using temporal convolutional networks (TCNs)~\cite{Steinmetz2021EfficientCompression}, and gray-box models based on DRC implementation structures~\cite{Wright2022Grey-boxCompression}.
While these attempts investigated various aspects of the topic, there is still room to improve the performance.
The output generated by some models still exhibits artifacts, and some best-performing models either rely on hard-coded components, are non-causal, or require a larger number of neural network parameters.
The need for a model with greater objective accuracy and perceptual quality that is causal, parameter efficient, and real-time capable remains.

In 2021, Gu et al. proposed S4, which implements an infinite impulse response (IIR) in the state-space form for long-sequence modeling~\cite{Gu2021EfficientlySpaces}.
It has proven powerful because it can have an arbitrarily long receptive field.
It can also preserve state information between samples or buffers to process arbitrarily long sequences.
Because of this, S4 seems promising for improving a model's performance in modeling DRCs.

This work proposes a model that uses S4 layers to characterize an analog DRC, namely the Teletronix LA-2A compressor.
This work introduced SSM to model an analog non-linear audio effect, exploring the effectiveness of using SSM in VA modeling.
Various experiments are conducted to evaluate our model's objective and subjective performance and real-time evaluation capabilities. 
The proposed model provides roughly the same quality as previous deep-learning models but with a causal formulation and fewer parameters.
It can also perform real-time inference given a specific audio buffer size, which is feasible in an audio production scenario.

\section{Background} \label{sec:bg}

\subsection{Structured State Space Sequence Model (S4)}

S4 is a neural network layer that implements an IIR system in state-space form. 
An $N^\text{th}$-order discrete SSM mapping of a mono signal to another mono signal can be expressed as follows, where $u$ is the input signal, $x$ is the intermediate signal, $y$ is the output signal, and $\mathbf{A} (N \times N), \mathbf{B} (N \times 1), \mathbf{C} (1 \times N), \mathbf{D} (1 \times 1)$ are state-space matrices expressing linear mappings.

\begin{align}
    x[t] & = \mathbf{A}x[t-1] + \mathbf{B}u[t] \\
    y[t] & = \mathbf{C}x[t]   + \mathbf{D}u[t]
\end{align}

Finite impulse response (FIR) systems like convolutional neural networks (CNNs) have finite-length impulse responses.
Since S4 is an IIR system, its impulse response is infinite, and its impulse response decay and receptive field are arbitrarily long.
S4 layers are also parameter efficient, given that for filters with a similar effect, IIR systems require fewer parameters.
There is an even more parameter-efficient S4 variant called S4D, with a diagonalized matrix $\mathbf{A}$~\cite{Gupta2022DiagonalSpaces}.

Inside an S4 layer, SSM matrices are complex-valued, allowing them to efficiently generate an impulse response that is as long as the input sequence using some mathematical techniques.
The input sequence is filtered using Fast Fourier transforms, producing the output sequence and state information.
S4 can process data sequences with state information preserved.
It can calculate the internal state at the end of one segment, allowing the next buffer to be computed without discontinuities.
For a very long sequence, one can input the entire sequence directly or section the sequences into small buffers and pass the state information.
The entire recursive process is causal.

\subsection{Feature-wise Linear Modulation}

Analog audio devices usually feature external controls that modify these devices' operation.
Digital emulations should also capture this behavior. 
To model DRC controls such as gain reduction, feature-wise linear modulation (FiLM) layers can be used~\cite{Perez2018FiLM:Layer}, following the approach of some other VA modeling research~\cite{Steinmetz2021EfficientCompression}.
Given an external information vector, FiLM first converts it into two vectors, $\gamma$ and $\beta$, using a multi-layer perceptron (MLP).
The output $y$ is given by $y=\gamma\odot~x~+~\beta$, where $x$ is the input and $\odot$ is element-wise vector multiplication.
FiLM layers enable adaptation of the model's behavior as a function of external controls.

\section{Methods} \label{sec:methods}

\subsection{Proposed Model}

\begin{figure}
    \centering
    \includegraphics[width=0.45\textwidth]{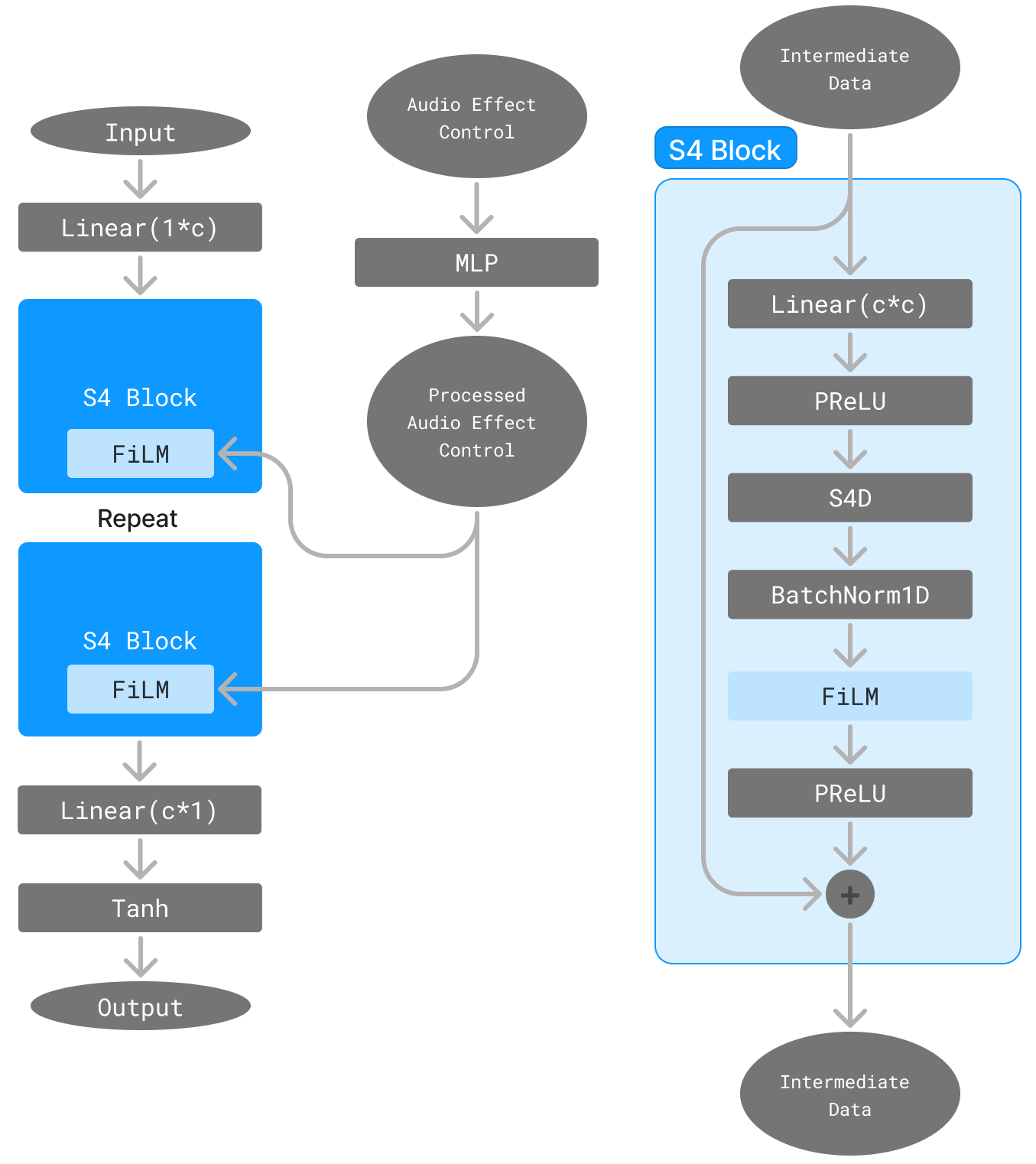}
    \caption{The proposed S4 model. It mainly comprises a stack of S4 blocks, where the S4 layer models the temporal dimension, the linear layer models the channel dimension, the FiLM layer applies external controls, and PReLU layers apply non-linearities.}
    \label{fig:s4-model}
\end{figure}

Our S4-based model is illustrated in Fig. \ref{fig:s4-model}.
At first, the input is expanded into $c$ data channels.
Unlike how CNN layers are conventionally implemented in DL, S4 layers are not combined or mixed across data channels.
Instead, every linear layer applies affine transformations on the data's channel dimension, thus creating a mix or combination of channel data on a frame-by-frame basis.

The design is essentially a chain of S4 blocks, illustrated at the right of Fig. \ref{fig:s4-model}.
Each block consists of a linear layer that mixes audio channels from $c$ to $c$, a PReLU layer to introduce non-linearity, an S4D layer, a BatchNorm1D layer, a FiLM layer to introduce audio effect controls, another PReLU layer, and a residual connection.
All S4 layers are S4D layers~\cite{Gupta2022DiagonalSpaces}. The
BatchNorm1D layers treat the channel dimension as the feature dimension.
An independent MLP processes external audio effect controls once to create control-based information before processing audio.
This control information is fed to all FiLM layers identically.
Each FiLM layer has its own independent MLP to convert control information to $\gamma$ and $\beta$ and apply them to the data's channel dimension.

After the stack of S4 blocks, the final linear layer contracts audio channels from $c$ to \num{1}.
The final $\tanh$ layer softly limits the output data to $\pm 1.0$.

\subsection{Experiment} \label{sec:method}

We tested our model with four S4 blocks, \num{16} or \num{32} inner audio channels, and fourth or eighth S4D SSM order.
There are four configurations in total.
Models are named in the format of \texttt{ssm-c*-f*}, where \texttt{c} means the inner audio channel number and \texttt{f} means S4 SSM order. 
Our training code and audio samples are available online~\footnote{\url{https://int0thewind.github.io/s4drc/}}.

The SignalTrain dataset~\cite{Colburn2020SignalTrainDataset} is used to train, validate, and test those models.
The SignalTrain dataset comprises audio input and output data processed by the Teletronix LA-2A compressor (LA-2A)
with different gain reductions and compressing/limiting switch values.
All audio data are mono sampled at \SI{44.1}{\kilo\Hz}.
There are \SI{87540}{\second} training data.
Phase inversion is the only data augmentation technique applied with a probability of \num{0.5}.
To accommodate the SignalTrain dataset, the model takes audio waveforms and audio effect controls as 32-bit floating point vectors. 
The input audio is mono with an amplitude range of $\pm 1.0$.

Models are trained using the SignalTrain dataset training split with batch size \num{32} in \num{60} epochs.
The training audio data are segmented into sections of length \num{65536} ($\approx$\SI{1.598}{\second} at \SI{44.1}{\kilo\Hz}).
No state information is preserved between audio buffers.
Each buffer is processed independently.
The learning rate is \num{0.001} and is reduced by a factor of 10 after ten epochs of no improvement in validation loss.
The optimizer is AdamW, with default function arguments from \texttt{PyTorch} (v2.0.0).
All S4D layer parameters' weight decays are set to zero.
The training loss function combines both time and frequency domain loss and is the sum of mean-averaged error (MAE) with multi-resolution STFT (multi-STFT) loss~\cite{Yamamoto2020ParallelSpectrogram}, with default function arguments from \texttt{auraloss} (v0.4.0)~\cite{Steinmetz2020Auraloss:PyTorch}.
Both parts are weighted equally.

Models are tested using the SignalTrain dataset testing split.
The testing audio data are segmented with length $2^{23}$ ($\approx$\SI{190.218}{\second} at \SI{44.1}{\kilo\Hz}) to test the model's long-term generalizability.
The entire buffer is fed into the model without slicing it.
When testing, MAE, mean-squared error (MSE), error-to-signal ratio (ESR) loss with a pre-emphasis filter of $H(z) = 1 - 0.85 z^{-1}$ plus DC loss (ESR+DC)~\cite{Wright2020PerceptualSystems}, multi-STFT loss with the same configuration in testing, loudness unit full scale (LUFS) difference with the ITU-R BS.1770 perceptual loudness recommendation, and Fr\'echet Audio Distance (FAD)~\cite{Kilgour2018FrechetAlgorithms} are evaluated.
MAE, MSE, and ESR+DC loss are time-domain criteria, and multi-STFT loss is a frequency-domain criterion.
ESR+DC loss may reflect the audio perceptual difference in the time domain, LUFS provides the loudness error, and FAD models perceptual similarity.
We took Steinmetz and Reiss' TCN and LSTM models~\cite{Steinmetz2021EfficientCompression} as the baseline, as S4 is closely related to TCN, and our model structure and training procedure are close to them.

\section{Result and Analysis} \label{sec:result-and-analysis}

\subsection{Objective Loss}

\input{tables/loss}

The test losses of our models and various baseline models are presented in Table \ref{tab:loss}.
Our \texttt{ssm-c32-f8} model has the best multi-STFT loss, and \texttt{ssm-c16-f8} has the best LUFS difference.
Other best metrics are from \texttt{TCN-300-N} and \texttt{LSTM-32}, yet we found that our \texttt{ssm-c32-f4} model has very close MAE and MSE, and our \texttt{ssm-c32-f8} has very close FAD.

We found the time domain losses of the \texttt{ssm-c16-f8} and \texttt{ssm-c32-f8} models to be greater than those of our other models.
Given that those higher loss values are from models with higher SSM filter orders, training a higher-order SSM might be more mathematically complicated.

We believe that our \texttt{ssm-c32-f4} model's performance is the most balanced. The 
\texttt{ssm-c32-f4} model has the best time-domain losses among all our models and outperforms all causal TCN models in all metrics.
It provides MAE and MSE performances that are close to those of \texttt{TCN-300-N}, which uses three times more model parameters than \texttt{ssm-c32-f4} and is not causal.
It also has a close FAD performance compared to \texttt{LSTM-32}, which cannot perform in real time and has higher time-domain and frequency-domain loss.
Although the \texttt{ssm-c32-f8} and \texttt{ssm-c16-f8} models have better multi-STFT loss and LUFS, their time-domain loss is much higher than that of the \texttt{ssm-c32-f4} model.
The \texttt{ssm-c32-f4} model has relatively good objective accuracy that is causal, parameter efficient, and real-time capable.

\subsection{Subjective Listening Study}

A multi-stimulus listening test similar to MUSHRA~\cite{Series2014MethodSystems} was conducted to further evaluate the model's performance.
The testing interface is webMUSHRA~\cite{Schoeffler2018WebMUSHRATests}.
It allows online assessment, and participants can instantaneously switch between clips to facilitate the comparison of minute differences.
Test participants score each audio clip based on the similarity to the reference and the effectiveness of the clip capturing the DRC's characteristics, with the range from \num{0} to \num{100}; the higher, the better.

\begin{figure}[h]
    \centering
    \includegraphics[width=0.4\textwidth]{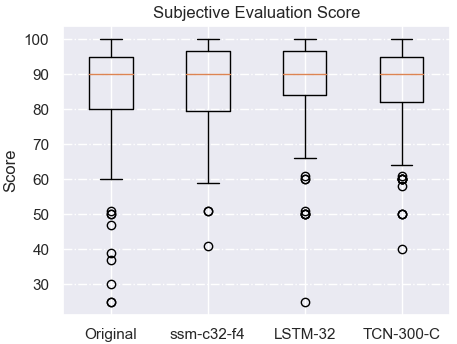}
    \caption{Subjective Evaluation Scores among all Clips}
    \label{fig:subjective}
\end{figure}

Eleven passages from the SignalTrain dataset testing split were included in the test, including strings, piano, guitar, and band clips.
Each audio clip has 10 seconds.
Three models were tested: \texttt{TCN-300-C} and \texttt{LSTM-32} from Steinmetz and Reiss~\cite{Steinmetz2021EfficientCompression} and our \texttt{ssm-c32-f4}.
We also include the original output clip in the test as a reference.
There are \num{17} valid responses.
The results are illustrated in Fig. \ref{fig:subjective}.

Generally, we found no immediately distinguishable results from the three models we tested.
All models show relatively the same median subjective scores and the same subjective score lower bounds, with our \texttt{ssm-c32-f4} model showing fewer outliers.
As most participants demonstrated, it is hard to distinguish differences between testing clips, making the scoring hard.
To conclude, there appears to be no significant difference in the ratings between our and baseline models.
The subjective listening study demonstrates a relatively close model performance between our S4 models and the previous TCN and LSTM models.

\subsection{Real-time Performance}

A real-time implementation requires that audio be processed incrementally in buffers to achieve a finite latency and that buffer computation time is less than the playback time.
To evaluate computation time, we processed multiple buffers using six sizes of 128 through 4096 samples with an Apple M1 Max CPU core (no GPU) and the \texttt{ssm-c32-f4} model (\SI{44.1}{\kilo\Hz} sampling rate), with state-passing.
Note that S4 layers can preserve state from one block to the next, eliminating discontinuities at block transitions.

\begin{figure}[h]
    \centering
    \includegraphics[width=0.4\textwidth]{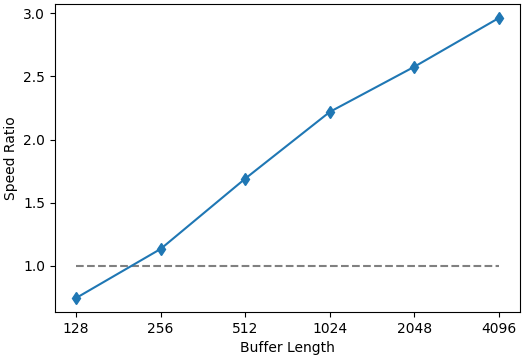}
    \caption{Speed Ratios in Different Buffer Lengths.}
    \label{fig:speed-ratio}
\end{figure}

We define ``speed ratio'' as the audio playback time divided by the buffer stream inference time in \texttt{PyTorch}'s inference mode.
A speed ratio higher than \num{1.0} means the inference speed is faster than the audio playback speed.
The speed ratio on those buffer streams is presented in Fig. \ref{fig:speed-ratio}.

The result shows that when the buffer size is greater than \num{256}, the inference speed is faster than real time.
Our implementation runs around three times as fast as real time using \num{4096} sample buffers.
This shows that our model can perform in real time and that the implementation is feasible in an audio production scenario.

Our S4 implementation produces impulse responses to process audio buffer-by-buffer rather than implementing SSM directly to process audio sample-by-sample.
The current S4 implementation might not portray the fastest real-time performance possible.
The learned compressor could also be implemented by applying the state space updates sample-by-sample, reducing latency.
We estimate this approach requires around 10M FLOPs per sample, so a single core capable of 5 GFLOPS should run faster than real time and perhaps even faster than the block-based approach.

\section{Conclusion} \label{sec:conclusion}

We presented a model that uses S4 layers to model an analog DRC.
Our model's objective performance and subjective performance are close to those of previous models but with a causal formulation and smaller model parameter space.
Specifically, our model can perform in real time on a CPU core with a buffer size greater than \num{256}, which is easily feasible in an audio production scenario.
We showed that a model with SSM can efficiently emulate an analog non-linear audio effect with long temporal dependencies, ensuring causality and a small parameter space.
While, in principle, SSMs process data sequences sample by sample, the current S4 implementation can only process data buffer by buffer.
A promising goal of future work could be the utilization of state-space matrices to process audio data sample-by-sample to evaluate the fastest real-time performance.

\vfill\pagebreak
\newpage

\bibliographystyle{IEEEbib} 
\bibliography{references}
\label{sec:refs}

\end{document}

%% file: tables/loss.tex
\begin{table*}[h]
    \centering
    \scalebox{0.85}{
    \begin{tabular}{lr|llllll}
    \toprule
        Model        & Params & MAE                & MSE                & ESR+DC             & Multi-STFT          & LUFS               & FAD                \\ \midrule
        \texttt{ssm-c16-f4} & 8.2k   & \texttt{1.012E-02} & \texttt{3.206E-04} & \texttt{4.003E-01} & \texttt{6.160E-01}  & \texttt{6.249E-01} & \texttt{4.813E-02} \\ 
        \texttt{ssm-c16-f8} & 9.3k   & \texttt{1.142E-01} & \texttt{2.432E-02} & \texttt{3.685E+00} & \texttt{6.588E-01}  & \texttt{\textbf{3.518E-01}} & \texttt{4.318E-02} \\ 
        \texttt{ssm-c32-f4} & 16.9k  & \texttt{\textit{8.737E-03}} & \texttt{\textit{2.879E-04}} & \texttt{\textit{4.065E-01}} & \texttt{4.881E-01}  & \texttt{4.766E-01} & \texttt{3.921E-02} \\ 
        \texttt{ssm-c32-f8} & 18.9k  & \texttt{1.157E-01} & \texttt{2.503E-02} & \texttt{3.676E+00} & \texttt{\textbf{4.785E-01}}  & \texttt{4.502E-01} & \texttt{\textit{3.713E-02}} \\ \midrule
        \texttt{TCN-100-N}    & 26k    & \texttt{1.580E-02} & \texttt{5.580E-04} & \texttt{2.331E-01} & \texttt{7.980E-01}  & \texttt{1.155E+00} & \texttt{1.599E+00} \\ 
        \texttt{TCN-300-N}    & 51k    & \texttt{\textbf{7.660E-03}} & \texttt{\textbf{1.350E-04}} & \texttt{\textbf{2.913E-02}} & \texttt{6.160E-01}  & \texttt{6.020E-01} & \texttt{1.062E-01} \\ 
        \texttt{TCN-1000-N}   & 33k    & \texttt{1.200E-01} & \texttt{2.650E-02} & \texttt{-} & \texttt{7.690E-01}  & \texttt{9.340E-01} & \texttt{1.762E+00} \\ 
        \texttt{TCN-100-C}    & 26k    & \texttt{1.920E-02} & \texttt{1.390E-03} & \texttt{1.880E+00} & \texttt{7.840E-01}  & \texttt{1.225E+00} & \texttt{1.903E+00} \\ 
        \texttt{TCN-300-C}    & 51k    & \texttt{1.440E-02} & \texttt{1.140E-03} & \texttt{1.800E+00} & \texttt{6.200E-01}  & \texttt{7.610E-01} & \texttt{1.036E-01} \\ 
        \texttt{TCN-1000-C}   & 33k    & \texttt{1.170E-01} & \texttt{2.570E-02} & \texttt{3.150E+00} & \texttt{7.100E-01}  & \texttt{8.990E-01} & \texttt{1.959E+00} \\ 
        \texttt{LSTM-32}      & 5k     & \texttt{1.100E-01} & \texttt{2.290E-02} & \texttt{1.870E+00} & \texttt{\textit{5.650E-01}}  & \texttt{\textit{3.610E-01}} & \texttt{\textbf{2.741E-02}} \\ \bottomrule
    \end{tabular}
    }
    \caption{Our model's testing metrics and baseline models metrics from Steinmetz and Reiss \cite{Steinmetz2021EfficientCompression}. Our models are on the top. \texttt{c} means inner audio channel number. \texttt{f} means S4 SSM order. For TCN models, \texttt{N} means non-causal. \texttt{C} means causal. Bold numbers are the global best metrics for all models. Italic numbers are the local best.}
    \label{tab:loss}
\end{table*}